\documentclass[sigconf]{acmart}

\IfFileExists{upquote.sty}{\usepackage{upquote}}{}
\IfFileExists{microtype.sty}{
  \usepackage[]{microtype}
  \UseMicrotypeSet[protrusion]{basicmath} 
}{}
\makeatletter
\@ifundefined{KOMAClassName}{
  \IfFileExists{parskip.sty}{%
    \usepackage{parskip}
  }{
    \setlength{\parindent}{0pt}
    \setlength{\parskip}{6pt plus 2pt minus 1pt}}
}{
  \KOMAoptions{parskip=half}}
\makeatother

\PassOptionsToPackage{unicode}{hyperref}
\PassOptionsToPackage{hyphens}{url}
\PassOptionsToPackage{dvipsnames,svgnames,x11names}{xcolor}

\IfFileExists{bookmark.sty}{\usepackage{bookmark}}{\usepackage{hyperref}}

\usepackage{color}
\usepackage{fancyvrb}

\DefineVerbatimEnvironment{Highlighting}{Verbatim}{commandchars=\\\{\}}
\newenvironment{Shaded}{}{}

\newcommand{\BuiltInTok}[1]{\textcolor[rgb]{0.65,0.15,0.64}{#1}}

\newcommand{\CommentTok}[1]{\textcolor[rgb]{0.63,0.63,0.65}{\textit{#1}}}

\newcommand{\ControlFlowTok}[1]{\textcolor[rgb]{0.65,0.15,0.64}{#1}}

\newcommand{\DecValTok}[1]{\textcolor[rgb]{0.60,0.41,0.00}{#1}}

\newcommand{\FloatTok}[1]{\textcolor[rgb]{0.60,0.41,0.00}{#1}}

\newcommand{\ImportTok}[1]{\textcolor[rgb]{0.31,0.63,0.31}{#1}}

\newcommand{\KeywordTok}[1]{\textcolor[rgb]{0.65,0.15,0.64}{#1}}
\newcommand{\NormalTok}[1]{\textcolor[rgb]{0.22,0.23,0.26}{#1}}
\newcommand{\OperatorTok}[1]{\textcolor[rgb]{0.65,0.15,0.64}{#1}}

\newcommand{\VariableTok}[1]{\textcolor[rgb]{0.89,0.34,0.29}{#1}}

\providecommand{\tightlist}{%
  \setlength{\itemsep}{0pt}\setlength{\parskip}{0pt}}\usepackage{longtable,booktabs,array}
\usepackage{calc} 
\usepackage{etoolbox}
\makeatletter
\patchcmd\longtable{\par}{\if@noskipsec\mbox{}\fi\par}{}{}
\makeatother
\IfFileExists{footnotehyper.sty}{\usepackage{footnotehyper}}{\usepackage{footnote}}
\makesavenoteenv{longtable}
\usepackage{graphicx}
\makeatletter
\newsavebox\pandoc@box
\newcommand*\pandocbounded[1]{
  \sbox\pandoc@box{#1}%
  \Gscale@div\@tempa{\textheight}{\dimexpr\ht\pandoc@box+\dp\pandoc@box\relax}%
  \Gscale@div\@tempb{\linewidth}{\wd\pandoc@box}%
  \ifdim\@tempb\p@<\@tempa\p@\let\@tempa\@tempb\fi
  \ifdim\@tempa\p@<\p@\scalebox{\@tempa}{\usebox\pandoc@box}%
  \else\usebox{\pandoc@box}%
  \fi%
}
\def\fps@figure{htbp}
\makeatother

\AtBeginEnvironment{longtable}{\tiny}
\usepackage{multirow}
\usepackage{bbm}
\definecolor{mypink}{RGB}{219, 48, 122}
\makeatletter
\@ifpackageloaded{caption}{}{\usepackage{caption}}
\AtBeginDocument{%
\ifdefined\contentsname
  \renewcommand*\contentsname{Table of contents}
\else
  \newcommand\contentsname{Table of contents}
\fi
\ifdefined\listfigurename
  \renewcommand*\listfigurename{List of Figures}
\else
  \newcommand\listfigurename{List of Figures}
\fi
\ifdefined\listtablename
  \renewcommand*\listtablename{List of Tables}
\else
  \newcommand\listtablename{List of Tables}
\fi
\ifdefined\figurename
  \renewcommand*\figurename{Figure}
\else
  \newcommand\figurename{Figure}
\fi
\ifdefined\tablename
  \renewcommand*\tablename{Table}
\else
  \newcommand\tablename{Table}
\fi
}
\@ifpackageloaded{float}{}{\usepackage{float}}
\floatstyle{ruled}
\@ifundefined{c@chapter}{\newfloat{codelisting}{h}{lop}}{\newfloat{codelisting}{h}{lop}[chapter]}
\floatname{codelisting}{Listing}

\makeatother
\makeatletter
\@ifpackageloaded{caption}{}{\usepackage{caption}}
\@ifpackageloaded{subcaption}{}{\usepackage{subcaption}}
\makeatother
\makeatletter
\@ifpackageloaded{tcolorbox}{}{\usepackage[skins,breakable]{tcolorbox}}
\makeatother
\makeatletter
\@ifundefined{shadecolor}{\definecolor{shadecolor}{rgb}{.97, .97, .97}}{}
\makeatother
\makeatletter
\@ifundefined{codebgcolor}{\definecolor{codebgcolor}{HTML}{F5F5F5}}{}
\makeatother
\makeatletter
\ifdefined\Shaded\renewenvironment{Shaded}{\begin{tcolorbox}[sharp corners, colback={codebgcolor}, boxrule=0pt, enhanced, frame hidden, breakable]}{\end{tcolorbox}}\fi
\makeatother

\hypersetup{
  pdftitle={Shallow AutoEncoding Recommender with Cold Start Handling via Side Features},
  pdfauthor={Edward DongBo Cui; Lu Zhang; William Ping-hsun Lee},
  colorlinks=true,
  linkcolor={blue},
  filecolor={Maroon},
  citecolor={Blue},
  urlcolor={red},
  pdfcreator={LaTeX via pandoc, via quarto}}

\AtBeginDocument{%
  }

\setcopyright{acmcopyright}
\copyrightyear{2025}
\acmYear{2025}
\acmDOI{XXXXXXX.XXXXXXX}

\acmConference[Conference acronym 'XX]{Conference}{XXXX, 2025}{XXXX}
\acmPrice{15.00}
\acmISBN{978-1-4503-XXXX-X/18/06}

\begin{document}

\title{Shallow AutoEncoding Recommender with Cold Start Handling via
Side Features}


  \author{Edward DongBo Cui}
  
            \affiliation{%
                  \institution{DotDash Meredith}
                                  \city{New York}
                                  \country{USA}
                      }
        \author{Lu Zhang}
  
            \affiliation{%
                  \institution{NBC Universal Peacock, equal
contributions}
                                  \city{Los Angeles}
                                  \country{USA}
                      }
        \author{William Ping-hsun Lee}
  
            \affiliation{%
                  \institution{NBC Universal Peacock, equal
contributions}
                                  \city{New York}
                                  \country{USA}
                      }

\begin{abstract}
User and item cold starts present significant challenges in industrial
applications of recommendation systems. Supplementing user-item
interaction data with metadata is a common solution, but often at the
cost of introducing additional biases. In this work, we introduce an
augmented EASE model that seamlessly integrates both user and item side
information to address these cold start issues. Our straightforward,
autoencoder-based method produces a closed-form solution that leverages
rich content signals for cold items while refining user representations
in data-sparse environments. Importantly, our method strikes a balance
by effectively recommending cold start items and handling cold start
users without incurring extra bias, and it maintains strong performance
in warm settings. Experimental results demonstrate improved
recommendation accuracy and robustness compared to previous
collaborative filtering approaches. Moreover, our model serves as a
strong baseline for future comparative studies.    
\end{abstract}



\maketitle

\setlength{\parskip}{-0.1pt}

\section{Introduction}\label{introduction}

The Embarrassing Shallow Autoencoder (EASE)
\citep{steckEmbarrassinglyShallowAutoencoders2019} is a
neighborhood-based collaborative filtering technique designed for top‑k
candidate generation in recommendation systems. Its straightforward
design, closed‑form solution, and robust performance have made it a
widely adopted baseline model in recommendation system research. Much
like other collaborative filtering approaches, the EASE model depends on
historical user-item interactions to uncover similarities between users
and items. This sole reliance on behavioral data means that such models
are not naturally equipped to handle the cold start problem. In the case
of the user cold start problem, new users have no interaction history,
which makes it challenging to accurately predict their preferences.
Conversely, the item cold start problem arises when new items have not
yet received any user feedback, making it hard to assess their quality
or categorize them appropriately. Both issues are common in real-world
recommendation systems, and effectively addressing them is essential for
enhancing recommendation accuracy and overall user experience in
personalized web applications.

User and item cold start problems are prevalent in the industrial
applications of recommendation systems and they may not occur
independently within individual applications. The present study has the
following contributions: It extends the previously developed conditional
autoencoding framework on recommendation. It further develops a
systematic approach to handle both user and item cold start problems
within the same modeling framework. Specifically, we leverage user and
item side information in a set of EASE-based autoencoding models to
enhance personalization experiences for newly onboarded users and
promote diversity of the recommendation by directing users' attention to
items that are yet to be discovered. The methodology we develop is
simple to implement given the existence of closed-form solution.
Experiments have shown that it can achieve better performance than
existing modeling frameworks that handle user and item cold start
problems simultaneously. We call our set of models from this methodology
\textbf{FEASE}, or \textbf{\emph{featurized}-EASE}.

The paper is organized as follows: in Section~\ref{sec-related-work}, we
first review a set of previous research related to neighborhood-based
and collaborative filtering-based approaches including EASE. Then in
Section~\ref{sec-methodology}, we describe the unified cold-start
methodology we developed, leveraging user and item side information.
Finally, in Section~\ref{sec-experimental-results}, we examine the
results of our methodology in a series of comparative studies across
various datasets, benchmarking it against several alternative
recommender models.

\section{Related Work}\label{sec-related-work}

In the current work, we focus on a class of models that can be viewed as
both an autoencoder in deep learning and a neighborhood-based model in
classic collaborative filtering approaches.

\subsection{EASE and SLIM}\label{ease-and-slim}

EASE \citep{steckEmbarrassinglyShallowAutoencoders2019} (backronym of
``Embarrassingly Shallow AutoEncoder'') is a neighborhood-based
auto-encoding recommendation model. Given a user-item interaction matrix
for \(N\) users and \(M\) items, \(X \in \mathbb{R}^{N \times M}\)
(though it is common to see EASE being applied to implicit feedback
data, where \(X_{ij} \in \{0, 1\}\)), the EASE model has the following
form

\begin{equation}\phantomsection\label{eq-ease-formulation}{\hat{X} = X B}\end{equation}

where

\begin{itemize}
\tightlist
\item
  \(B\) is a square weight matrix, with
  \(B \in \mathbb{R}^{M \times M}\) and constraint
  \(\text{diag}(B) = 0\).
\item
  \(\hat{X}\) is the dense estimated interaction score matrix.
\end{itemize}

The EASE model is optimized by minimizing the following regression loss
with respect to \(B\)

\begin{equation}\phantomsection\label{eq-ease-loss-func}{L(B) = \|X - XB\|_F^2 + \lambda \|B\|_F^2 + 2 \gamma^\top \text{diag}(B)}\end{equation}

where \(\|\cdot \|_F^2\) is the Frobenius norm of the matrix. Then \(B\)
has a closed-form solution as follows

\begin{equation}\phantomsection\label{eq-ease-closed-form-solution}{\hat{B} = I - P \cdot \text{diagMat}(\vec{1} \oslash \text{diag}(P))}\end{equation}

where

\begin{itemize}
\tightlist
\item
  \(P = (X^\top X - \lambda I)^{-1}\).
\item
  \(\oslash\) is the element-wise division between vector inputs.
\item
  \(\text{diagMat}(\cdot)\) is the diagonal matrix formed by the vector
  input.
\end{itemize}

EASE is closely related to SLIM \citep{ningSLIMSparseLinear2011}
(\textbf{S}parse \textbf{LI}near \textbf{M}ethods) model given by

\begin{equation}\phantomsection\label{eq-slim-formulation}{\hat{X} = XW}\end{equation}

with the constraints \(W \ge 0\) and \(\text{diag}(W) = 0\). It is
optimized by minimizing the following regression loss

\begin{equation}\phantomsection\label{eq-slim-loss-func}{L(W) = \frac{1}{2} \|X - XW\|_F^2 + \frac{\beta}{2}\|W\|_F^2 + \lambda \|W\|_1}\end{equation}

where \(\|\cdot\|_1\) is the L1 norm. In comparison, EASE model drops
the L1 regularization term as well as the non-negative value constraint
on the weight matrix \(W\) (i.e.~\(W \ge 0\)) and found the
auto-encoding model work equally well in recommendation tasks.

\subsection{AutoEncoder}\label{autoencoder}

The class of models that we are focusing on in the present study is
generally related to auto-encoder based collaborative filtering models.
AutoRec \citep{sedhainAutoRecAutoencodersMeet2015} is among the first to
view the collaborative filtering problem as a reconstruction problem and
introduced deep learning concepts in recommender systems. AutoRec is a
two-layer neural network model with the form

\[h(X; \theta) = g(V \cdot f(WX + b) + c)\]

where

\begin{itemize}
\tightlist
\item
  \(X \in \mathbb{R}^{N \times M}\) is the user-item interaction matrix.
  Note that this is specifically for the AutoRec-I or item-based
  formulation. AutoRec can be applied for both explicit and implicit
  feedback data.
\item
  \(W \in \mathbb{R}^{M \times K}\) is the weight matrix for the hidden
  layer, with \(K\) being the number of hidden layer neurons (a.k.a
  hidden dimensions).
\item
  \(V \in \mathbb{R}^{K \times M}\) is the weight matrix for the output
  layer.
\item
  \(b\) and \(c\) are the bias terms of the hidden and output layers,
  respectively.
\item
  \(f\) and \(g\) are activation functions, e.g.~sigmoid function, for
  the hidden and output layers, respectively.
\item
  \(\theta\) is the set of parameters of the model,
  i.e.~\(\theta \in \{W, V, b, c\}\).
\end{itemize}

Both EASE and SLIM models can be viewed as modifications of the AutoRec
model, where

\begin{enumerate}
\def\labelenumi{\arabic{enumi}.}
\tightlist
\item
  The bias terms are dropped.
\item
  Activations functions \(f\) and \(g\) are identity functions (i.e.~no
  activations).
\item
  The weight terms are combined
  \(B = W \cdot V = \mathbb{R}^{M \times K} \cdot \mathbb{R}^{K \times M} = \mathbb{R}^{M \times M}\).
\item
  Additional constraints and regularizations are being added to
  regularize the combined weight matrix to reduce overfitting.
\end{enumerate}

In addition to AutoRec, other more sophisticated forms of auto-encoding
recommendation models were also developed, such as Multinomial
Variational AutoEncoder (Mult-VAE)
\citep{liangVariationalAutoencodersCollaborative2018} and Collaborative
Deep Denoising AutoEncoder (CDDAE)
\citep{zhaoCollaborativeDeepDenoising2016} for more efficient and
optimal learning of the user preference reconstruction.

\subsection{Neighborhood-based Collaborative
Filtering}\label{neighborhood-based-collaborative-filtering}

Both EASE and SLIM are also closely related to neighborhood-based
collaborative filtering, e.g.~ItemKNN
\citep{deshpandeItembasedTopRecommendation2004a}. In neighborhood-based
models, a square matrix \(S \in \mathbb{R}^{M \times M}\) is computed to
store similarity scores between each item pair. Then to provide
recommendations for each user, the scores are aggregated across the
corresponding rows of the square matrix, forming a final vector of
length \(M\) with each element corresponding to a score of the
recommended item. This score aggregation scheme specifically corresponds
to EASE or SLIM when applied to implicit-feedback data, where the
user-item interaction input is \(X_i \in \{0, 1\}^{1 \times M}\). Then
the score aggregation is simply

\[X_i \cdot S = \hat{X}_i\]

where \(\hat{X}_i\) is also row vectors of length \(M\). Notice that
this is the same formulation as EASE
(Equation~\ref{eq-ease-formulation}) or SLIM
(Equation~\ref{eq-slim-formulation}). For explicit feedback, scores are
weighted by the historical feedback/ratings before being combined.
ItemKNN uses either cosine-similarity between item vectors of user
purchases (views, engagement, etc.) or modified conditional
probabilities between pairs of co-purchased (or co-viewed, co-engaged,
co-occurrence, etc.) items to construct the square matrix \(S\). The
weight matrices \(B\) in EASE and \(W\) in SLIM can also be interpreted
as a similarity matrix between items. More specifically, the closed-form
solution given by EASE reveals that weight matrix \(B\) is the
regularized inverse of the Graham matrix of the user-item interaction
data, i.e.~\(G = X^T X\).

\subsection{EASE Model with Item Side
Information}\label{ease-model-with-item-side-information}

The EASE model was adapted to incorporate item side-information, as seen
in the collective-EASE model (CEASE)
\citep{jeunenClosedFormModelsCollaborative2020}. This is similar to the
collective-SLIM model \citep{ningSparseLinearMethods2012}. The modified
loss for CEASE is given by

\begin{equation}\phantomsection\label{eq-collective-ease-loss-func}{L(B) = \|X - XB\|_F^2 + \alpha \|T - TB\|_2^F + \lambda \|B\|_F^2}\end{equation}

subject to \(\text{diag}(B) = 0\). The model has a closed-form solution
given by

\[\hat{B} = I - P \cdot \text{diagMat}(\vec{1} \oslash \text{diag}(P))\]

which is the same as the closed form solution seen in the original EASE
model, but with

\begin{itemize}
\tightlist
\item
  \(P = (\tilde{X}^\top Q \tilde{X} + \lambda I)^{-1}\)
\item
  \(\tilde{X} = \begin{bmatrix} X \\ T \end{bmatrix}\)
\item
  \(Q \in \mathbb{R}^{ (N+L) \times (N+L) }\) is a diagonal weight
  matrix that regularizes the importance of each item tag. If all users
  have the weight of 1 and all tags have a constant weight of \(\alpha\)
  as in the original loss function
  (Equation~\ref{eq-collective-ease-loss-func}), then
\end{itemize}

\[Q = \begin{bmatrix} 1 & 0 & 0 & ... & ... & ... & ... & ... \\ 0 & 1 & 0 & ... & ... & ... & ... & ... \\0 & 0 & 1 & ... & ... & ... & ... & ... \\ ... & ... & ... & ... & ... & ... & ... & ... \\ ... & ... & ... & ... & \alpha & 0 & 0 & ... \\ ... & ... & ... & ... & 0 & \alpha & 0 & ... \\ ... & ... & ... & ... & 0 & 0 & \alpha & ... \\ ... & ... & ... & ... & ... & ... & ... & ... \\ \end{bmatrix}\]

Equivalently, we can also redefine
\(\tilde{X} = \begin{bmatrix} X \\ \sqrt{\alpha} T \end{bmatrix}\).

Meta-data Alignment for cold-start Recommendation (MARec)
\citep{monteilMARecMetadataAlignment2024a} is another cold start
approach that leverages item side information. It can be applied to not
only the EASE model but also other autoencoder based such as SLIM and
variational autoencoders. When applied to a EASE backbone model, it has
a closed form solution similar to EASE.

Although these approaches can handle the item cold-start problem to some
extent, previous research have not specifically applied these models and
evaluated their performance when both item and user cold start problems
simultaneously exist.

\section{Methodology}\label{sec-methodology}

In this section, we introduce our methodology to handle the user and
item cold start problems simultaneously, with minimal impact on the
performance of recommendation for warm users and items.

\subsection{User Cold Start}\label{user-cold-start}

We further extend the formulation given by collective-EASE and develop
our AutoEncoder model that handles user cold-start problem for implicit
feedback recommendations by simultaneously leveraging user and item side
features.

We construct our input data \(Z\) as a sparse matrix as follows

\[Z = \begin{bmatrix} X & \beta U \\ \alpha T  & \mathbf{0} \end{bmatrix}\]

where

\begin{itemize}
\tightlist
\item
  \(Z \in \mathbb{R}^{(N+L) \times (M+K)}\).
\item
  \(X \in \{0, 1\}^{N \times M}\) is the user-item interaction matrix
  (\(N\) users, \(M\) items) with implicit feedbacks.
\item
  \(T \in \{0, 1\}^{L \times M}\) is the tag-item indicator matrix
  (\(L\) tags) for all items.
\item
  \(U \in \{0, 1\}^{N \times K}\) is the user-attribute indicator matrix
  (\(K\) attributes) for all users.
\item
  \(\alpha\) and \(\beta\) are constant weights for item tags and user
  attributes, respectively.
\end{itemize}

We can then define a model like EASE

\[\hat{Z} = ZS\]

subject to \(\text{diag}(S) = 0\), with the learning objective

\[L(S) = \|Z - ZS\|_F^2 + \lambda \|S\|_F^2 + 2 \gamma^\top \cdot \text{diag}(S)\]

We denote this formulation of the FEASE model as \textbf{FEASE-U}
(i.e.~featurized-EASE with user cold start). This is identical to the
original EASE formulation, therefore, the weight matrix estimate
\(\hat{S}\) has a closed-form solution

\[\hat{S} = I - P \cdot \text{diagMat}\Big(\vec{1} \oslash \text{diag}(P) \Big)\]
where \(P = (Z^\top Z + \lambda I)^{-1}\).

\subsection{Item Cold Start}\label{item-cold-start}

As we will see in the experiment results
(Section~\ref{sec-item-cold-start-handling}), although we have
incorporated item side information in the model, the augmented EASE
formulation above still cannot solve the item cold start problem. With
additional analyses, we found that cold items in matrix \(B\) in EASE
are assigned with zeros and matrix \(S\) in FEASE-U are assigned with
random scores close to 0 (see Section~\ref{sec-warm-cold-item-scores}),
which are uninformative for the task of recommendation. This is simply
because there is no user-item interaction data to allow the model to
learn and assign a useful score on the cold items. To see why the score
for cold items are uninformative, we can examine the EASE model through
a Bayesian reformulation. Let

\[p(X | B, \sigma^2) = \mathcal{N}(X; XB, \sigma^2 I) \]

Then by Bayes' rule,

\[p(X | B, \sigma^2) = \frac{p(B |X, \sigma^2) p(X)}{p(B)}\]

Rearrange the terms

\[p(B|X, \sigma^2) \propto p(X|B, \sigma^2) p(B)\]

Then estimating \(B\) is the same as maximizing the posterior
\(p(B|X, \sigma^2)\). However, for cold items, the likelihood
\(p(X|B, \sigma^2\)) will be a constant (i.e.~always 0-scored for all
users regardless of the value in \(B\)), that is,

\[p(B | X, \sigma^2) \propto p(B)\]

The value in \(B\) will depend only on the prior \(p(B)\), which may be
a constant or a randomly initialized value unrelated to user
preferences. Therefore, one way to mitigate this is to provide a better
prior value of \(B\) or a default score on the cold items. A simple
strategy is to leverage item-to-item similarity based on content
metadata. We can then formulate a new optimization objective similar to
EASE as follows

\begin{equation}\phantomsection\label{eq-fease-i-prior}{\begin{aligned} L(B) = \|X - XB\|_F^2 + \lambda \|B\|_F^2 + \delta \|B - R\|_F^2 \\ + 2 \gamma^\top \cdot \text{diagMat}(B) \end{aligned}}\end{equation}

where \(R \in \mathbb{R}^{M \times M}\) is an item-to-item similarity
score matrix and \(\delta\) is a regularization weight. This encourages
the learned matrix \(B\) to fall back onto \(R\) if no data in \(X\) can
inform the value in \(B\). Carefully tuning \(\delta\) can balance the
trade-offs between the joint optimization of the loss terms
\(\|X - XB\|^2_F\) and \(\|B - R\|^2_F\), letting both user-item
interaction and content similarity to contribute to the final item
scores in \(B\). We denote this formulation of our model as
\textbf{FEASE-I-Prior} (featurized-EASE with item cold start, jointly
optimized with a prior of \(B\)). The above loss formulation in
Equation~\ref{eq-fease-i-prior} also has a closed-form solution. Taking
derivative on both sides with respect to \(B\),

\[\begin{aligned} \frac{\partial L(B)}{\partial B} = -2 X^\top X + 2X^\top XB + 2 \lambda B \\+ 2 \delta(B - R) + 2 \text{diagMat}(\gamma) = 0 \end{aligned}\]

Rearranging the terms,

\[\big(X^\top X + (\lambda + \delta) I\big)B + \text{diagMat}(\gamma) = X^\top X + \delta R\]

Let \(P = \big(X^\top X + (\lambda + \delta)I \big)^{-1}\). Solving for
\(B\),

\[\hat{B} = P(X^\top X+ \delta R) - P \cdot \text{diagMat}(\gamma)\]

To enforce zero diagonal constraint in \(B\),
i.e.~\(\text{diag}(B) = 0\),

\[\begin{aligned} \text{diag}(\hat{B}) = \text{diag}\big(P(X^\top X + \delta R) \big) \\- \text{diag}(P \cdot \text{diagMat}(\gamma) = 0 \end{aligned}\]

Let \(d = \text{diag}\big(P(X^\top X+ \delta R)\big)\) and
\(p = \text{diag}(P)\), the above can be simplified to

\[d - p \odot \gamma = 0 \implies \gamma = \frac{d}{p}\]

where \(\odot\) stands for element-wise multiplication, and \(\oslash\)
stands for element-wise division. The final solution for \(B\) becomes

\begin{equation}\phantomsection\label{eq-fease-i-prior-solution}{\begin{aligned} \hat{B} &= P(X^\top X + \delta R) \\ &- P \cdot \text{diagMat}\big(\text{diag}\big(P(X^\top X+ \delta R)\big) \oslash \text{diag}(P) \big) \end{aligned}}\end{equation}

One potential caveat with the formulation in
Equation~\ref{eq-fease-i-prior} is that we cannot limit the optimization
to the cold items only. Given the user-item behavior data is more
informative of the user preferences than simple content similarity in
most applications, incorporating the additional \(\|B - R\|^2_F\) term
for warm items can reduce the performance of the recommendation on these
items that have significant amount of signals from user interactions.
One way to mitigate this is to decouple the calculation of warm and cold
item scores in the final matrix \(B\), and only optimize the
\(\|B - R\|_F^2\) term for the cold items. We can define a
mask/indicator matrix \(\mathbf{1}_C \in \{0, 1\}^{M \times M}\) such
that,

\[(\mathbf{1}_C)_{ij} = \begin{cases} 1 \quad \text{if either } i,j \in \text{\{cold items\}} \\ 0 \quad \text{otherwise} \end{cases}\]

Then the loss function is changed to

\begin{equation}\phantomsection\label{eq-fease-i-prior-masked}{\begin{aligned} L(B) &= \|X - XB\|_F^2 + \lambda \|B\|_F^2 \\ &+ \delta \|\mathbf{1}_C \odot (B - R)\|_F^2 + 2 \gamma^\top \cdot \text{diagMat}(B) \end{aligned}}\end{equation}

But we can also skip this optimization entirely and instead use a simple
heuristic as follows to achieve the same goal:

\begin{itemize}
\tightlist
\item
  For warm items, use scores obtained from the EASE model.
\item
  For cold items, use scores obtained from content similarity weighted
  by an additional tunable scale factor \(\delta\).
\end{itemize}

This way, cold items are independently modeled and scored, while
minimally impacting the recommendation quality of the warm items. We
call this formulation of our model \textbf{FEASE-I} (featurized-EASE
with item cold start). The above heuristic can be implemented easily by
replacing the values of the cold items in the matrix \(B\):

\begin{itemize}
\tightlist
\item
  Rescaling: Min-max scale \(R\) to the minimum and maximum values of
  \(B\). Then multiply the rescaled \(R\) with an additional scale
  factor \(\delta\). This can give cold items additional boosts to help
  them show up in the top-k recommendations.
\item
  Row-wise replacement: Replace the rows of cold items in \(B\) with the
  corresponding rows in \(R\).
\item
  Column-wise replacement: Replace the columns of the cold items in
  \(B\) with corresponding columns in \(R\).
\end{itemize}

\subsection{Simultaneous User and Item Cold Start
Handling}\label{simultaneous-user-and-item-cold-start-handling}

We can further combine FEASE-U and FEASE-I (or FEASE-I-Prior) together
to formulate a new model that can handle user and item cold start
problems simultaneously within a single framework. We denote this as the
\textbf{FEASE} model, which is described as follows

Let input data \(Z\) be a sparse matrix constructed as in FEASE-U

\[Z = \begin{bmatrix} X & \beta U \\ \alpha T  & \mathbf{0} \end{bmatrix}\]

and a zero-padded content similarity matrix

\[R' = \begin{bmatrix} R & \mathbf{0} \\ \mathbf{0} & \mathbf{0} \end{bmatrix}\]

Similar to the FEASE-I-Prior formulation, \textbf{FEASE-Prior} can be
solved by minimizing the following loss function

\[\begin{aligned} L(S) &= \|Z - ZS\|_F^2 + \lambda \|S\|_F^2 \\ & + \delta \|S - R'\|_F^2 + 2 \gamma^\top \cdot \text{diag}(S) \end{aligned}\]

with a closed-form solution as

\[\begin{aligned} \hat{S} &= P(Z^\top Z + \delta R') \\ &- P \cdot \text{diagMat}\big(\text{diag}\big(P(Z^\top Z + \delta R')\big) \oslash \text{diag}(P) \big) \end{aligned}\]

where \(P = \big(Z^\top Z + (\lambda + \delta)I \big)^{-1}\).

For the full \textbf{FEASE} formulation (i.e.~the decoupled model), to
solve for \(S\), we first solve the FEASE-U problem

\[\hat{S} = I - P \cdot \text{diagMat}\Big(\vec{1} \oslash \text{diag}(P) \Big)\]

where \(P = (Z^\top Z + \lambda I)^{-1}\). Then we apply the heuristics
for FEASE-I to create the final weight matrix \(\hat{S}\) by replacing
the cold-items with the values in \(R'\). Note that when min-max
normalizing, we will only compute the statistics on the \(M \times M\)
block part in \(\hat{S}\) (i.e.~\texttt{S\_hat{[}:M,\ :M{]}}).

\subsection{Similarity Model}\label{similarity-model}

Since the similarity matrix \(R\) serves as the prior weight matrix of
\(B\), it needs to be predictive of the user's preferences to some
extent, even though not modeling directly from the user preference data.
To keep our model construction as straightforward as possible, we
generate item representations using TF-IDF, leveraging item tags,
descriptions, and other relevant metadata, and then compute similarities
via cosine similarity scores on these TF-IDF embedding vectors.
Investigating alternative embedding approaches---such as those based on
large language models (LLMs) or other similarity scoring methods---is
beyond the scope of this study and will be explored in future work.

\subsection{Implementation in Python}\label{implementation-in-python}

Below is a code snippet on how to implement FEASE-I-Prior using Python
(FEASE-Prior requires only changes in the input), given the Gram matrix
input \(G \in \mathbb{R}^{(M+K) \times (M+K)}\) and the padded
similarity matrix \(R'\).

\scriptsize

\begin{Shaded}
\begin{Highlighting}[numbers=left,,]
\ImportTok{import}\NormalTok{ numpy }\ImportTok{as}\NormalTok{ np}
\KeywordTok{def}\NormalTok{ compute\_weight\_matrix\_with\_prior(}
\NormalTok{   G: np.ndarray,}
\NormalTok{   R\_prime: np.ndarray,}
\NormalTok{   lambda\_reg: }\BuiltInTok{float}\NormalTok{,}
\NormalTok{   delta\_reg: }\BuiltInTok{float}\NormalTok{, }
\NormalTok{) }\OperatorTok{{-}\textgreater{}}\NormalTok{ np.ndarray:}
    \CommentTok{"""Item cold start handling of (F)EASE }
\CommentTok{    model by jointly optimizing a prior R\textquotesingle{}."""}
\NormalTok{    P }\OperatorTok{=}\NormalTok{ G }\OperatorTok{*} \DecValTok{1} \CommentTok{\# copy}
\NormalTok{    diag\_ind }\OperatorTok{=}\NormalTok{ np.diag\_indices(P.shape[}\DecValTok{0}\NormalTok{])}
\NormalTok{    P[diag\_ind] }\OperatorTok{+=}\NormalTok{ (lambda\_reg }\OperatorTok{+}\NormalTok{ delta\_reg)}
\NormalTok{    P }\OperatorTok{=}\NormalTok{ np.linalg.inv(P) }
\NormalTok{    G }\OperatorTok{=}\NormalTok{ G }\OperatorTok{+}\NormalTok{ delta\_reg }\OperatorTok{*}\NormalTok{ R\_prime}
\NormalTok{    S }\OperatorTok{=}\NormalTok{ P }\OperatorTok{@}\NormalTok{ G }\CommentTok{\# unconstrained}
\NormalTok{    S }\OperatorTok{=}\NormalTok{ S }\OperatorTok{{-}}\NormalTok{ P }\OperatorTok{@}\NormalTok{ np.diag(np.diag(S) }\OperatorTok{/}\NormalTok{ np.diag(P))}
\NormalTok{    S }\OperatorTok{=}\NormalTok{ S.astype(np.float32)}
\NormalTok{    np.fill\_diagonal(S, }\DecValTok{0}\NormalTok{)}
    \ControlFlowTok{return}\NormalTok{ S}
\end{Highlighting}
\end{Shaded}

\normalsize

The below code snippet illustrates how to merge together the matrix
\(B\) from the original EASE formulation (or the \(M \times M\) block
part in \(\hat{S}\)) and the content-similarity based matrix \(R\), in
the decoupled FEASE-I and the full FEASE formulation.

\scriptsize

\begin{Shaded}
\begin{Highlighting}[numbers=left,,]
\KeywordTok{def}\NormalTok{ merge\_R\_B\_matrices(}
\NormalTok{    R: np.ndarray, }
\NormalTok{    B: np.ndarray, }
\NormalTok{    is\_cold\_item: np.ndarray, }
\NormalTok{    weight: }\BuiltInTok{float} \OperatorTok{=} \FloatTok{1.0}\NormalTok{,}
\NormalTok{) }\OperatorTok{{-}\textgreater{}}\NormalTok{ np.ndarray:}
    \CommentTok{\# Rescale R\textquotesingle{}s stats to B\textquotesingle{}s stats}
\NormalTok{    B\_min, B\_max }\OperatorTok{=}\NormalTok{ np.}\BuiltInTok{min}\NormalTok{(B), np.}\BuiltInTok{max}\NormalTok{(B)}
\NormalTok{    R\_min, R\_max }\OperatorTok{=}\NormalTok{ np.}\BuiltInTok{min}\NormalTok{(R), np.}\BuiltInTok{max}\NormalTok{(R)}
\NormalTok{    R }\OperatorTok{=}\NormalTok{ (R }\OperatorTok{{-}}\NormalTok{ R\_min) }\OperatorTok{/}\NormalTok{ (R\_max }\OperatorTok{{-}}\NormalTok{ R\_min) }
\NormalTok{    R }\OperatorTok{=}\NormalTok{ R }\OperatorTok{*}\NormalTok{ (B\_max }\OperatorTok{{-}}\NormalTok{ B\_min) }\OperatorTok{+}\NormalTok{ B\_min}
    \CommentTok{\# give additional preference to R}
\NormalTok{    R }\OperatorTok{=}\NormalTok{ R }\OperatorTok{*}\NormalTok{ weight}
    \CommentTok{\# Taking care of cold start item rows}
\NormalTok{    B }\OperatorTok{=}\NormalTok{ np.where(is\_cold\_item[:, }\VariableTok{None}\NormalTok{], R, B)}
    \CommentTok{\# Also take care of the columns: making warm }
    \CommentTok{\# items also recommend cold items}
\NormalTok{    B }\OperatorTok{=}\NormalTok{ np.where(is\_cold\_item[}\VariableTok{None}\NormalTok{, :], R, B)}
    \ControlFlowTok{return}\NormalTok{ B}
\end{Highlighting}
\end{Shaded}

\normalsize

The above two functions can then be used to implement the FEASE models,
using standard Python numerical libraries such as
NumPy\citep{harrisArrayProgrammingNumPy2020}.

\section{Experimental Results}\label{sec-experimental-results}

\subsection{Datasets}\label{datasets}

We evaluated our models along with various baseline models against
datasets commonly used in the recommendation literature, as shown in
Table~\ref{tbl-dataset-description}: Netflix
\citep{netflixNetflixPrizeData2006}, MovieLens
\citep{harperMovieLensDatasetsHistory2016}, AmazonBooks
\citep{niJustifyingRecommendationsUsing2019}. When splitting the
datasets into train-validation-test sets, we generated cold users by
leaving certain users only in the validation and test sets. For those
datasets that lacks specific user side features, we also augmented the
data with additional user side features such as number of ratings
(bucketized into distinct categories) and day of the week of interaction
to help handle the user cold start problem. Each user will get a single
combination of user side features. Therefore, for the engineered
features such as day of the week, if a user had interactions on multiple
days of the week, then the interactions were treated as if they are from
different users. Similarly, to generate cold items, we leave certain
items only in the validation and test sets, making sure they never show
up in the training set. We then leverage as much side information about
the items available within the dataset to handle the item cold start
problem.

\subsection{Baseline Models}\label{baseline-models}

We compare our FEASE model against baseline approaches, which include
models from the EASE family as well as other techniques designed to
address user and/or item cold start challenges. The models include

\begin{itemize}
\tightlist
\item
  \textbf{Popularity}: top-k recommendation by popularity of the items
  (denoted as ``Popularity''). In addition to recommending by the
  overall item popularity, we can also make the recommendation by
  computing the popularity within each user segment (i.e.~combinations
  of user feature values), which we denote as (``Popularity(seg)'').
\item
  \textbf{EASE} \citep{steckEmbarrassinglyShallowAutoencoders2019}: this
  is the baseline model that FEASE and other family members of our
  methodology are derived from.
\item
  \textbf{CEASE} \citep{jeunenClosedFormModelsCollaborative2020}: or
  collective-EASE model that leverages only the item side features. We
  mainly evaluate its capability of handling the item cold start
  problem.
\item
  \textbf{MARec} \citep{monteilMARecMetadataAlignment2024a}: MARec
  leverages item side information, which achieved state of the art
  performance on various datasets to handle cold start. We used the
  closed-form solution as described in Equation 9 of the original paper.
  We also obtained the code implementation from the original authors.
\item
  \textbf{Factorization Machine (FM)}
  \citep{rendleFactorizationMachines2010}: this is a strong baseline for
  simultaneous user and item cold start handling.
\item
  \textbf{DropoutNet} \citep{volkovsDropoutNetAddressingCold2017}: this
  is another strong baseline that augments on Matrix Factorization for
  simultaneous user and item cold start handling. We train the Matrix
  Factorization model using Bayesian Personalized Ranking (BPR)
  \citep{rendleBPRBayesianPersonalized2009}.
\end{itemize}

\begin{table}

\caption{\label{tbl-dataset-description}Summary of test datasets with
cold start.}

\centering{

\begin{center}
\scriptsize
\renewcommand{\arraystretch}{1.2}
\begin{tabular}{lcccc}
\hline
\textbf{Dataset} & \textbf{\# Cold Users} & \textbf{\# Cold Items} & \textbf{User Features} & \textbf{Item Features} \\
\hline
Netflix & 19,388 & 1,004 & \parbox[t]{1.2cm}{{\fontsize{6}{8}\selectfont DayofWeek\\\#Ratings}} & \parbox[t]{1.2cm}{{\fontsize{6}{8}\selectfont Tags\\Year\\Description}} \\
\hline
MovieLens & 11,474 & 1,103 & \parbox[t]{1.2cm}{{\fontsize{6}{8}\selectfont DayofWeek\\HourofDay\\\#Ratings}} & \parbox[t]{1.2cm}{{\fontsize{6}{8}\selectfont Tags}} \\
\hline
Amazon Books & 1,480 & 1,000 & \parbox[t]{1.2cm}{{\fontsize{6}{8}\selectfont DayofWeek\\\#Ratings\\\#Reviews}} & \parbox[t]{1.2cm}{{\fontsize{6}{8}\selectfont Tags}} \\
\hline
\end{tabular}
\end{center}

}

\end{table}%

\subsection{Evaluation Metrics}\label{evaluation-metrics}

We evaluated our recommender models using standard top-k metrics in the
literature \citep{steckEmbarrassinglyShallowAutoencoders2019}, namely,
Hit Ratio (HR), Recall, Normalized Discounted Cumulative Gain (NDCG),
and Effective Catalog Size (ECS)
\citep{gomez-uribeNetflixRecommenderSystem2015}. Let \(u \in U\) be the
user, \(i \in I\) be the item, \(R_u^{(K)}\) be the top-k
recommendation, \(T_u\) is the target label based on user-item
interaction history, and \(\mathbbm{1}\) be the indicator operator.

Hit Ratio is defined as

\[\text{HR}@K = \frac{1}{|U|}\sum_{u \in U} \mathbbm{1}\{R_u^{(K)} \cap T_u\} \]

Recall is defined as

\[\text{Recall}@K\frac{1}{|U|} \sum_{u \in U} \frac{|R_u^{(K)}\cap T_u|} {|R_u |}\]

NDCG is defined as

\[\text{NDCG}@K = \frac{1}{|U|} \sum_{u \in U} \frac{\text{DCG@K}}{IDCG@K}\]

where \(\text{DCG}@K = \sum_{i=1}^K \frac{1}{\log_2(i+1)}\),
\(\text{IDCG}@K = \sum_{i=1}^{|T_u|} \frac{1}{\log_2{(i+1)}}\), given
equal relative importance of each label and recommendation.

ECS is defined as

\[\text{ECS}@K = 2 \Big( \sum_{r=1}^N p_r \cdot r  \Big) - 1\]

where \(p_r\) is the normalized fraction of item at rank \(r\) being
recommended in top-k recommendations, with \(p_r > p_{r+1}\), and
\(r = 1, ..., N\). ECS is used as a measurement of recommendation
diversity, which is another important metric for recommendation quality
that is less studied in the literature.

To measure the effectiveness of recommending cold items that aligns with
the user's interests, we adapt the Hit Ratio metric for item cold start
(ColdHR) as

\[\text{ColdHR}@K = \frac{1}{|U'|} \sum_{u \in U'} \mathbbm{1}\{R_u^{(K)} \cap T_u^{(C)} \}\]
where \(U'\) is the set of test users who have cold item in their
user-item interaction, and \(T_u^{(C)}\) is the set of cold items that
user \(u\) interacted with.

\subsection{Overall Performance}\label{overall-performance}

Table~\ref{tbl-overall-performance} shows the performance of models on
the test splits of various datasets, including both warm and cold users
and items. The FEASE model family performs generally well comparing to
other baselines, suggesting the effectiveness of incorporating user
and/or item side information on the task of top-k recommendations. Note
that the collective-EASE (or CEASE) model performs equally or marginally
better than the original EASE model, suggesting that incorporating item
side information as part of the input data matrix can be helpful, but
not as significant comparing to when incorporating user side
information, as seen in the FEASE-U model. A significant improvement in
recommendation accuracy is possibly made by item popularity, as both
popularity based methods contribute a significant portion of the model
performance. The FEASE-I model may appear to perform worse than even the
baseline EASE model in the current results, but this is because we tuned
the hyperparameters so that the model can recommend cold items with a
small sacrifice on warm item recommendation. Consequently, FEASE-I model
has the better recommendation diversity than other EASE model family
members (see Table~\ref{tbl-item-diversity-performance}), suggesting a
trade-off between recommendation accuracy and diversity.

\begin{table}

\caption{\label{tbl-overall-performance}Comparisons of model performance
on the entire dataset. Bold text indicates the highest performing
metric.}

\centering{

\centering
\begingroup
\fontsize{6}{8}\selectfont
\renewcommand{\arraystretch}{1.2}

\begin{minipage}{\linewidth}
\subcaption{Netflix}
\centering
\begin{tabular}{p{1.4cm}p{0.5cm}p{0.5cm}p{0.75cm}|p{0.5cm}p{0.5cm}p{0.75cm}}
\hline & \multicolumn{3}{c|}{@20} & \multicolumn{3}{c}{@50} \\ \hline
\textbf{Model} & \textbf{HR} & \textbf{Recall} & \textbf{NDCG} & \textbf{HR} &  \textbf{Recall} &  \textbf{NDCG} \\
\hline
Popularity & 0.1985 & 0.0457 & 0.0803 & 0.3371 & 0.0950 & 0.1071 \\
\hline
Popularity(seg) & 0.2106 & 0.0545 & 0.0867 & 0.3527 & 0.1063 & 0.1145 \\
\hline
MFBPR & 0.2707 & 0.0785 & 0.1126 & 0.4287 & 0.1458 & 0.1431 \\
\hline
DropoutNet & 0.2915 & 0.0832 & 0.1222 & 0.4584 & 0.1560 & 0.1546 \\
\hline
FM & 0.3111 & 0.0960 & 0.1320 & 0.4818 & 0.1731 & 0.1648 \\
\hline
EASE & 0.4230 & 0.1426 & 0.2033 & 0.5598 & 0.2226 & 0.2261 \\
\hline
CEASE & 0.4230 & 0.1426 & 0.2033 & 0.5598 & 0.2226 & 0.2261 \\
\hline
MARec & 0.4248 & 0.1434 & 0.2044 & 0.5614 & 0.2240 & 0.2271 \\
\hline
FEASE-U & \textbf{0.4342} & \textbf{0.1450} & \textbf{0.2080} & 0.5764 & 0.2270 & \textbf{0.2320} \\
\hline
FEASE-I & 0.4004 & 0.1302 & 0.1762 & 0.5577 & 0.2212 & 0.2060 \\
\hline
FEASE & 0.4209 & 0.1378 & 0.1848 & \textbf{0.5792} & \textbf{0.2299} & 0.2147 \\
\hline
\end{tabular}
\end{minipage}

\par\bigskip 

\begin{minipage}{\linewidth}
\subcaption{MovieLens-20M}
\centering
\begin{tabular}{p{1.4cm}p{0.5cm}p{0.5cm}p{0.75cm}|p{0.5cm}p{0.5cm}p{0.75cm}}
\hline & \multicolumn{3}{c|}{@20} & \multicolumn{3}{c}{@50} \\ \hline
\textbf{Model} & \textbf{HR} & \textbf{Recall} & \textbf{NDCG} & \textbf{HR} &  \textbf{Recall} &  \textbf{NDCG} \\
\hline
Popularity & 0.2535 & 0.0705 & 0.1154 & 0.3711 & 0.1202 & 0.1372 \\
\hline
Popularity(seg) & 0.2597 & 0.0733 & 0.1166 & 0.3753 & 0.1232 & 0.1394 \\
\hline
MFBPR & 0.2295 & 0.0734 & 0.0963 & 0.3557 & 0.1338 & 0.1205 \\
\hline
DropoutNet & 0.3012 & 0.0928 & 0.1362 & 0.4331 & 0.1594 & 0.1606 \\
\hline
FM & 0.2830 & 0.0973 & 0.1269 &  0.4059 & 0.1677 & 0.1493 \\
\hline
EASE & 0.4569 & 0.1908 & 0.2383 & 0.5684 & 0.2819 & 0.2542 \\
\hline
CEASE & 0.4569 & 0.1908 & 0.2383 & 0.5685 & 0.2820 & 0.2542 \\
\hline
MARec & 0.4549 & 0.1907 & 0.2376 & 0.5645 & 0.2802 & 0.2531 \\
\hline
FEASE-U & \textbf{0.4965} & \textbf{0.1956} & \textbf{0.2559} & \textbf{0.6222} & 0.2897 & 0.2746 \\
\hline
FEASE-I & 0.4433 & 0.1829 & 0.2241 & 0.5626 & \textbf{0.2971} & \textbf{0.2945} \\
\hline
FEASE & 0.4892 & 0.1921 & 0.2461 & 0.6203 & 0.2898 & 0.2669 \\
\hline
\end{tabular}
\end{minipage}

\par\bigskip 

\begin{minipage}{\linewidth}
\subcaption{Amazon Books}
\centering
\begin{tabular}{p{1.4cm}p{0.5cm}p{0.5cm}p{0.75cm}|p{0.5cm}p{0.5cm}p{0.75cm}}
\hline & \multicolumn{3}{c|}{@20} & \multicolumn{3}{c}{@50} \\ \hline
\textbf{Model} & \textbf{HR} & \textbf{Recall} & \textbf{NDCG} & \textbf{HR} &  \textbf{Recall} &  \textbf{NDCG} \\
\hline
Popularity & 0.0299 & 0.0178 & 0.0118 & 0.0590 & 0.0353 & 0.0176 \\
\hline
Popularity(seg) & 0.0320 & 0.0189 & 0.0128 & 0.0608 & 0.0364 & 0.0187 \\
\hline
MFBPR & 0.1226 & 0.0787 & 0.0510 & 0.2129 & 0.1410 & 0.0686 \\
\hline
DropoutNet & 0.1114  & 0.0701 & 0.0460 & 0.1973 & 0.1279 & 0.0627 \\
\hline
FM & 0.0488 & 0.0331 & 0.0143 & 0.1360 & 0.0901 & 0.0316 \\
\hline
EASE & 0.2012 & 0.1311 & 0.1076 & 0.2710 & 0.1818 & 0.1204 \\
\hline
CEASE & 0.2012 & 0.1313 & 0.1064 & 0.2745 & 0.1844 & 0.1199 \\
\hline
MARec & 0.2011 & 0.1310 & 0.1076 & 0.2711 & 0.1819 & 0.1204 \\
\hline
FEASE-U & \textbf{0.2052} & \textbf{0.1325} & \textbf{0.1078} & \textbf{0.2821} & \textbf{0.1872} & \textbf{0.1220} \\
\hline
FEASE-I & 0.1871 & 0.1205 & 0.0876 & 0.2703 & 0.1811 & 0.1034 \\
\hline
FEASE & 0.2049 & 0.1288 & 0.0970 & 0.2811 & 0.1865 & 0.1122 \\
\hline
\end{tabular}
\end{minipage}


\endgroup

}

\end{table}%

\subsection{Results for Cold Users}\label{results-for-cold-users}

Table~\ref{tbl-user-cold-start-performance} shows the performance of
recommendation on cold users only. It is consistent with the expectation
that models incorporating user side information can handle user cold
start, i.e.~make user-segment level recommendations that can align with
user's interests. Notice that Popularity-based models, DropoutNet and/or
Factorization Machine perform better than other models across various
datasets, followed by FEASE and FEASE-U models. However, these models
gained the better capability of handling cold users by largely
sacrificing its performance on warm users, as seen in
Table~\ref{tbl-overall-performance}. Therefore, the benefit of FEASE
model is that it can handle user cold start without impacting warm user
performance at all. In fact, its performance can be better for those
models solely focuses on warm users, such as Matrix Factorization and
EASE model.

\begin{table}

\caption{\label{tbl-user-cold-start-performance}Comparisons of model
performance on cold users. Bold text indicates the highest performing
metric.}

\centering{

\centering
\begingroup
\fontsize{6}{8}\selectfont
\renewcommand{\arraystretch}{1.2}

\begin{minipage}{\linewidth}
\subcaption{Netflix}
\centering
\begin{tabular}{p{1.4cm}p{0.5cm}p{0.5cm}p{0.75cm}|p{0.5cm}p{0.5cm}p{0.75cm}}
\hline & \multicolumn{3}{c|}{@20} & \multicolumn{3}{c}{@50} \\ \hline
\textbf{Model} & \textbf{HR} & \textbf{Recall} & \textbf{NDCG} & \textbf{HR} &  \textbf{Recall} &  \textbf{NDCG} \\
\hline
Popularity & 0.3454 & 0.0627 & \textbf{0.1586} & 0.4801 & 0.1115 & 0.1860 \\
\hline
Popularity(seg) & 0.3155 & 0.0707 & 0.1295 & 0.4849 & 0.1279 & 0.1665 \\ 
\hline
MFBPR & 0.0173 & 0.0139 & 0.0064 & 0.0378 & 0.0288 & 0.0105 \\ 
\hline
DropoutNet & \textbf{0.3625} &\textbf{0.0727} & 0.1566 & \textbf{0.5027} & \textbf{0.1286} & \textbf{0.1871} \\ 
\hline
FM & 0.2364 & 0.0530 & 0.0992 & 0.4247 & 0.1073 & 0.1386 \\ 
\hline
EASE & 0.0322 & 0.0025 & 0.0138 & 0.0516 & 0.0035 & 0.0175 \\
\hline
CEASE & 0.0322 & 0.0025 & 0.0138 & 0.0516 & 0.0035 & 0.0175 \\ 
\hline
MARec & 0.0322 & 0.0025 & 0.0138 & 0.0516 & 0.0035 & 0.0175 \\ 
\hline
FEASE-U & 0.3107 & 0.0569 & 0.1297 & 0.4666 & 0.1037 & 0.1635 \\ 
\hline
FEASE-I & 0.0322 & 0.0025 & 0.0138 & 0.0516 & 0.0035 & 0.0175 \\
\hline
FEASE & 0.3107 & 0.0569 & 0.1297 & 0.4666 & 0.1037 & 0.1635 \\
\hline
\end{tabular}
\end{minipage}

\par\bigskip 

\begin{minipage}{\linewidth}
\subcaption{MovieLens-20M}
\centering
\begin{tabular}{p{1.4cm}p{0.5cm}p{0.5cm}p{0.75cm}|p{0.5cm}p{0.5cm}p{0.75cm}}
\hline & \multicolumn{3}{c|}{@20} & \multicolumn{3}{c}{@50} \\ \hline
\textbf{Model} & \textbf{HR} & \textbf{Recall} & \textbf{NDCG} & \textbf{HR} &  \textbf{Recall} &  \textbf{NDCG} \\
\hline
Popularity & 0.4675 & \textbf{0.0942} & \textbf{0.2384} & 0.5905 & \textbf{0.1332} & \textbf{0.2636} \\
\hline
Popularity(seg) & \textbf{0.4704} & 0.0828 & 0.2170 & \textbf{0.6022} & 0.1175 & 0.2453 \\
\hline
MFBPR & 0.0112 & 0.0028 & 0.0039 & 0.0284 & 0.0054 & 0.0073 \\
\hline
DropoutNet & 0.4555 & 0.0905 & 0.2277 & 0.5796 & 0.1304 & 0.2535 \\
\hline
FM & 0.0046 & 0.0027 & 0.0015 & 0.0110 & 0.0062 & 0.0028 \\
\hline
EASE & 0.0016 & 0.0001 & 0.0008 & 0.0040 & 0.0002 & 0.0013 \\
\hline
CEASE & 0.0016 & 0.0001 & 0.0008 & 0.0040 & 0.0002 & 0.0013 \\
\hline
MARec & 0.0016 & 0.0001 & 0.0008 & 0.0040 & 0.0002 & 0.0013 \\
\hline
FEASE-U & 0.4177 & 0.0621 & 0.1908 & 0.5552 & 0.0886 & 0.2182 \\
\hline
FEASE-I & 0.0016 & 0.0001 & 0.0008 & 0.0040 & 0.0002 & 0.0013 \\
\hline
FEASE & 0.4252 & 0.0644 & 0.1938 & 0.5573 & 0.0907 & 0.2201 \\
\hline
\end{tabular}
\end{minipage}

\par\bigskip 

\begin{minipage}{\linewidth}
\subcaption{Amazon Books}
\centering
\begin{tabular}{p{1.4cm}p{0.5cm}p{0.5cm}p{0.75cm}|p{0.5cm}p{0.5cm}p{0.75cm}}
\hline & \multicolumn{3}{c|}{@20} & \multicolumn{3}{c}{@50} \\ \hline
\textbf{Model} & \textbf{HR} & \textbf{Recall} & \textbf{NDCG} & \textbf{HR} &  \textbf{Recall} &  \textbf{NDCG} \\
\hline
Popularity & \textbf{0.0846} & 0.0272 & 0.0341 & \textbf{0.1591} & 0.0510 & \textbf{0.0487} \\
\hline
Popularity(seg) & 0.0732 & 0.0250 & 0.0305 & 0.1330 & 0.0480 & 0.0466 \\
\hline
MFBPR & 0.0267 & 0.0137 & 0.0109  & 0.0521 & 0.0287  & 0.0159 \\
\hline
DropoutNet &  0.0631 & 0.0211 & 0.0257 & 0.1116 & 0.0412 & 0.0350 \\
\hline
FM & 0.0840 & \textbf{0.0290} & \textbf{0.0347} & 0.1443 & \textbf{0.0543} & 0.0467 \\
\hline
EASE & 0.0034 & 0.0006 & 0.0011 & 0.0089 & 0.0019 & 0.0022 \\
\hline
CEASE & 0.0034 & 0.0006 & 0.0011 & 0.0089  & 0.0019 & 0.0022 \\
\hline
MARec & 0.0049 & 0.0010 & 0.0015 & 0.0110 & 0.0021 & 0.0027 \\
\hline
FEASE-U & 0.0815 & 0.0251 & 0.0320 & 0.1517  & 0.0502  & 0.0457 \\
\hline
FEASE-I & 0.0034 & 0.0006 & 0.0011 & 0.0089 & 0.0019 & 0.0022 \\
\hline
FEASE & 0.0797 & 0.0246 & 0.0310 & 0.1502 & 0.0493 & 0.0450 \\
\hline
\end{tabular}
\end{minipage}

\endgroup

}

\end{table}%

\subsection{Item Cold Start
Handling}\label{sec-item-cold-start-handling}

Table~\ref{tbl-item-diversity-performance} illustrates the model's
performance on the items diversity (as measured by ECS@K) and cold item
recommendation accuracy (as measured by ColdItemHR@K). In general, it is
expected that models incorporating item side information can handle item
cold start as well. However, both CEASE and FEASE-U incorporated item
side information as part of their optimization, but we see no gain in
item cold start performance comparing to the EASE model. This suggests
that special model design is needed to effectively use item side
information to solve the item cold start problem. Additionally,
Factorization Machine model is not always effective despite using item
side information as inputs during inference.

On the other hand, combining with the observation of
Table~\ref{tbl-overall-performance}, models with better recommendation
accuracy (e.g.~high Hit Ratio) in general has correspondingly lower item
diversity (i.e.~ECS). Therefore, there is a trade-off between item
diversity and accuracy in recommendation. Lastly, with proper tuning of
the prior matrix weights, FEASE models can outperform other models that
can handle the item cold start problem, while not significantly
impacting the performance on warm items.

\begin{table}

\caption{\label{tbl-item-diversity-performance}Comparison of model
performance on item diversity and cold item recommendations. Bold text
indicates the highest performing metric.}

\centering{

\centering
\begingroup
\fontsize{6}{8}\selectfont
\renewcommand{\arraystretch}{1.2}

\begin{minipage}{\linewidth}
\subcaption{Netflix}
\centering
\begin{tabular}{p{1.4cm}p{1.0cm}p{1.0cm}|p{1.0cm}p{1.0cm}}
\hline & \multicolumn{2}{c|}{@20} & \multicolumn{2}{c}{@50} \\ \hline
\textbf{Model} & \textbf{ColdHR} & \textbf{ECS} & \textbf{ColdHR} & \textbf{ECS} \\
\hline
Popularity         & 0.       & 23.3  & 0.      & 56.3 \\
\hline
Popularity(seg)    & 0.       & 44.9  & 0.      & 84.8 \\
\hline
MFBPR           & 0.       & \textbf{921.4} & 0. & \textbf{1029.3} \\
\hline
DropoutNet      & 0.02324  & 491.0 & 0.06486  & 611.5  \\
\hline
FM              & 0.00040  & 521.2 & 0.00101  & 633.2  \\
\hline
EASE            & 0.00004  & 587.3 & 0.00020  & 731.7  \\
\hline
CEASE           & 0.00004  & 588.0 & 0.00020  & 732.6  \\
\hline
MARec           & 0.00334  & 699.3 & 0.01345  & 841.9  \\
\hline
FEASE-U         & 0.       & 541.8 & 0.       & 662.5  \\
\hline
FEASE-I         & \textbf{0.07621}  & 611.5 & \textbf{0.09057} & 815.4  \\
\hline
FEASE           & 0.06240  & 577.9 & 0.06708  & 715.9  \\
\hline
\end{tabular}
\end{minipage}

\par\bigskip 

\begin{minipage}{\linewidth}
\subcaption{MovieLens-20M}
\centering
\begin{tabular}{p{1.4cm}p{1.0cm}p{1.0cm}|p{1.0cm}p{1.0cm}}
\hline & \multicolumn{2}{c|}{@20} & \multicolumn{2}{c}{@50} \\ \hline
\textbf{Model} & \textbf{ColdHR} & \textbf{ECS} & \textbf{ColdHR} & \textbf{ECS} \\
\hline
Popularity & 0. & 25.2 & 0. & 58.7 \\
\hline
Popularity(seg) & 0 & 37.2 & 0 & 87.0 \\
\hline
MFBPR & 0. & \textbf{2337.8} & 0. & \textbf{2472.5} \\
\hline
DropoutNet & 0.00110 & 559.6 & 0.00434 & 722.1 \\
\hline
FM &  0. & 655.8 & 0.00002 & 829.5 \\
\hline
EASE & 0. & 547.8 & 0.01048 & 754.6 \\
\hline
CEASE & 0. & 548.1 & 0. & 755.1 \\
\hline
MARec & 0.00469 & 607.2 & 0.02705 & 857.5 \\
\hline
FEASE-U & 0. & 505.7 & 0. & 694.1 \\
\hline
FEASE-I & \textbf{0.02910} & 580.4 & \textbf{0.06386} & 813.4 \\
\hline
FEASE & 0.02265 & 520.5 & 0.04013 & 716.9 \\
\hline
\end{tabular}
\end{minipage}

\par\bigskip 

\begin{minipage}{\linewidth}
\subcaption{Amazon Books}
\centering
\begin{tabular}{p{1.4cm}p{1.0cm}p{1.0cm}|p{1.0cm}p{1.0cm}}
\hline & \multicolumn{2}{c|}{@20} & \multicolumn{2}{c}{@50} \\ \hline
\textbf{Model} & \textbf{ColdHR} & \textbf{ECS} & \textbf{ColdHR} & \textbf{ECS} \\
\hline
Popularity & 0. & 20.2 & 0. & 50.4  \\
\hline
Popularity(seg) & 0. & 62.0 & 0. & 143.7 \\
\hline
MFBPR & 0. & 883.5 & 0. & 1472.9 \\
\hline
DropoutNet & 0. & 724.9 & 0. & 1280.9 \\
\hline
FM & 0. & \textbf{4651.9} & 0. & \textbf{3778.0} \\
\hline
EASE & 0.00011 &  2945.0& 0.00015 & 3775.9 \\
\hline
CEASE & 0.00011 & 2588.9 & 0.00015  & 3390.4 \\
\hline
MARec & 0.00103 & 2952.4 & 0.00349  & 3794.8 \\
\hline
FEASE-U & 0. & 2418.0 & 0. & 3134.9 \\
\hline
FEASE-I & \textbf{0.01173} & 2272.9 & \textbf{0.01346} & 3398.1 \\
\hline
FEASE & 0.00743 & 2546.7 & 0.00817 & 3301.5 \\
\hline
\end{tabular}
\end{minipage}

\endgroup

}

\end{table}%

\subsection{Difference in Scores Between Warm and Cold
items}\label{sec-warm-cold-item-scores}

\begin{figure}

\centering{

\pandocbounded{\includegraphics[keepaspectratio]{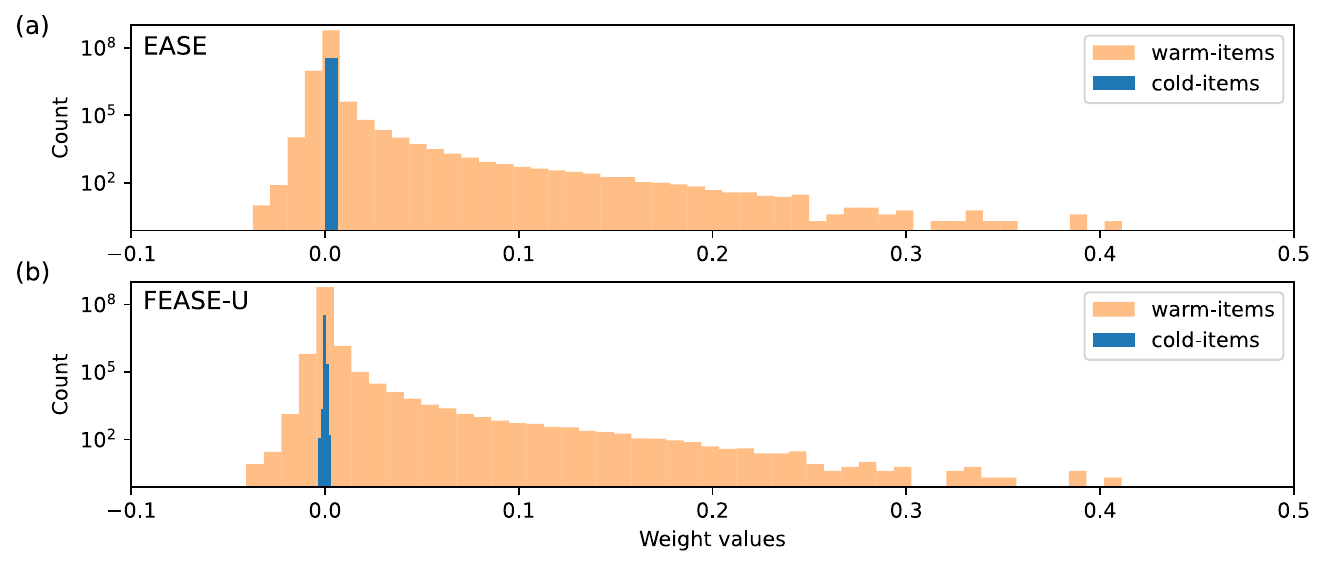}}

}

\caption{\label{fig-warm-cold-item-scores}Learned weights from EASE (a)
and FEASE-U (b) models on Netflix data.}

\end{figure}

To help explain why the EASE family models cannot handle items in their
recommendations, we evaluated the score distributions of the warm
vs.~cold items in the EASE family model.
Figure~\ref{fig-warm-cold-item-scores} illustrates the score
distributions of cold and warm items for both EASE
(Figure~\ref{fig-warm-cold-item-scores}a) and FEASE-U
(Figure~\ref{fig-warm-cold-item-scores}b) are near zero, which are in
contrast with a wide distribution of mostly positive scores of the warm
items. This result suggests that the cold items are not receiving proper
assignments of the weight scores that provide sufficient information on
user preferences. Recommendations of the cold items are therefore mostly
deprioritized by the model.

\subsection{Effects of Splitting Users by
Context}\label{effects-of-splitting-users-by-context}

In the Netflix dataset, we augmented the user-item interaction matrix
with user side features such as day of the week. This enables us to
treat the same user as distinct entities under different contexts
(e.g.~when interacting on different days), thereby capturing temporal
variations in user behaviors. We compare two variants of the EASE model:
one trained on these contextually split users (``EASE'', essentially
treating users under different context as separate users), and another
trained on the complete data for each user (denoted as ``EASE(full)'').
Both models are evaluated on the user-segmented data. As shown in
Table~\ref{tbl-ease-variant-performance}, the model trained on segmented
users performs better overall, even though neither variant effectively
addresses the user or item cold start problem. This finding underscores
the importance of train-test alignment in machine learning. This result
also illustrates that user side features may contain significant
underlying structure regarding user preferences; the differences are
substantial enough that simply treating interactions as if they come
from different users effectively captures their contextual preferences,
even without explicitly incorporating side information into the model.

\begin{table}

\caption{\label{tbl-ease-variant-performance}Comparison between two
variants of the EASE model with user-segmented vs.~unsegmented Netflix
training data: EASE vs.~EASE(full).}

\centering{

\centering
\begingroup
\fontsize{6}{8}\selectfont
\renewcommand{\arraystretch}{1.2}

\begin{minipage}{\linewidth}
\subcaption{Overall performance}
\centering
\begin{tabular}{p{1.3cm}p{0.5cm}p{0.5cm}p{0.75cm}|p{0.5cm}p{0.5cm}p{0.75cm}}
\hline & \multicolumn{3}{c|}{@20} & \multicolumn{3}{c}{@50} \\ \hline
\textbf{Model} & \textbf{HR} & \textbf{Recall} & \textbf{NDCG} & \textbf{HR} &  \textbf{Recall} &  \textbf{NDCG} \\
\hline
EASE & 0.4230 & 0.1426 & 0.2033 & 0.5598 & 0.2226 & 0.2261 \\
\hline
EASE(full) & 0.4148 & 0.1394 & 0.1973 & 0.5530 & 0.2187 & 0.2208 \\
\hline
\end{tabular}
\end{minipage}

\par\bigskip 

\begin{minipage}{\linewidth}
\subcaption{Cold user performance}
\centering
\begin{tabular}{p{1.3cm}p{0.5cm}p{0.5cm}p{0.75cm}|p{0.5cm}p{0.5cm}p{0.75cm}}
\hline & \multicolumn{3}{c|}{@20} & \multicolumn{3}{c}{@50} \\ \hline
\textbf{Model} & \textbf{HR} & \textbf{Recall} & \textbf{NDCG} & \textbf{HR} &  \textbf{Recall} &  \textbf{NDCG} \\
\hline
EASE & 0.0322 & 0.0025 & 0.0138 & 0.0516 & 0.0035 & 0.0175 \\
\hline
EASE(full) & 0.0322 & 0.0025 & 0.0138 & 0.0516 & 0.0035 & 0.0175 \\
\hline
\end{tabular}
\end{minipage}

\par\bigskip 

\begin{minipage}{\linewidth}
\subcaption{Item diversity and cold Item recommendation}
\centering
\begin{tabular}{p{1.3cm}p{1.0cm}p{1.0cm}|p{1.0cm}p{1.0cm}}
\hline & \multicolumn{2}{c|}{@20} & \multicolumn{2}{c}{@50} \\ \hline
\textbf{Model} & \textbf{ColdHR} & \textbf{ECS} & \textbf{ColdHR} & \textbf{ECS} \\
\hline
EASE            & 0.00004  & 587.3 & 0.00020  & 731.7  \\
\hline
EASE(full)      & 0.00004  & 650.1 & 0.00020  & 806.0  \\
\hline
\end{tabular}
\end{minipage}

\endgroup

}

\end{table}%

\subsection{Joint Optimization vs.~Cold Item Weight
Replacement}\label{joint-optimization-vs.-cold-item-weight-replacement}

\begin{figure}

\centering{

\pandocbounded{\includegraphics[keepaspectratio]{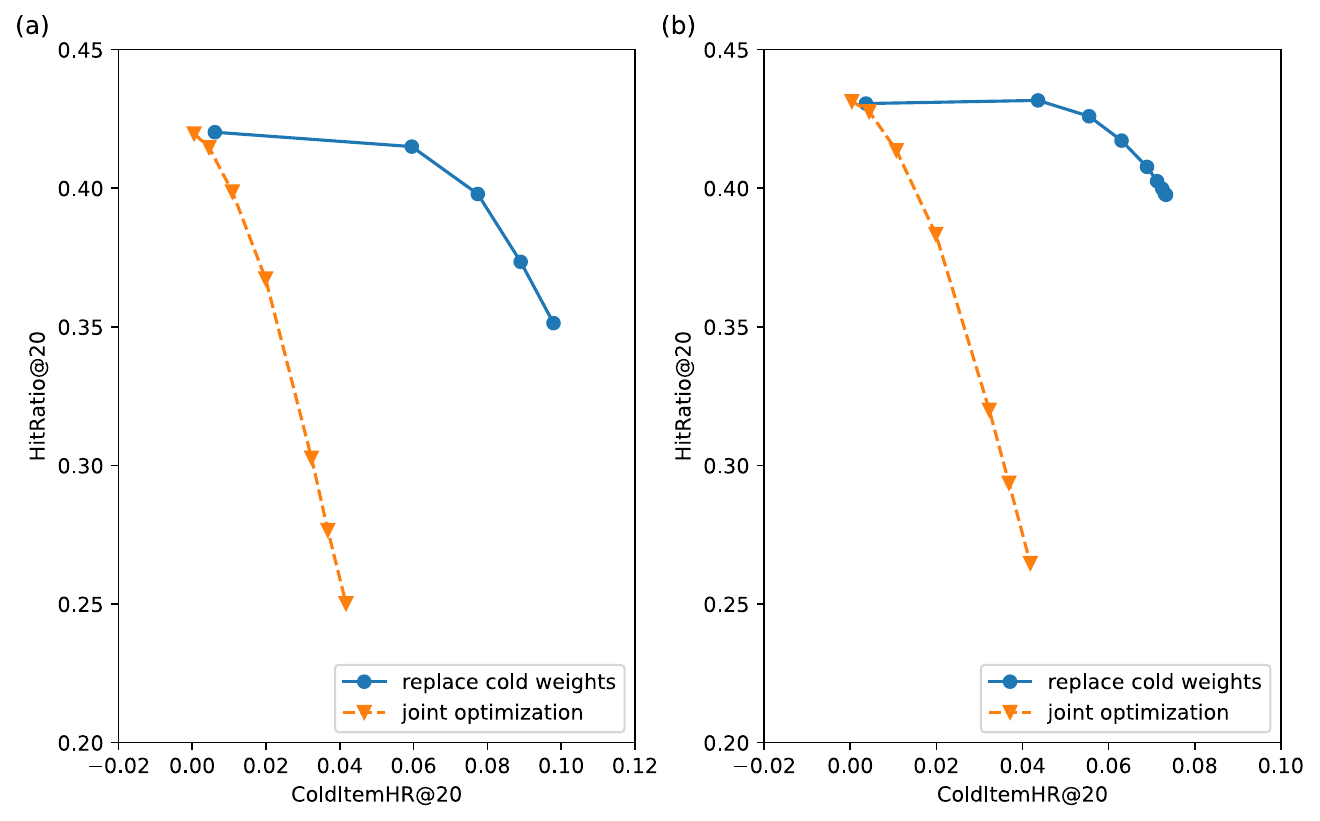}}

}

\caption{\label{fig-joint-vs-replacement}Model performance comparison
between joint optimization (i.e.~models with ``-Prior'' suffix) and cold
item weight replacement on Netflix data. a) FEASE-I (solid line)
vs.~FEASE-I-Prior (dashed line). b) FEASE (solid line) vs.~FEASE-Prior
(dashed line). Panel (a) shows FEASE-I (solid line) versus FEASE-I-Prior
(dashed line), and panel (b) compares FEASE (solid line) to FEASE-Prior
(dashed line). In both cases, as the weight \(\delta\) increases,
overall model performance (measured by HitRatio@20 on the y-axis)
declines, while the exposure of cold items (measured by ColdItemHR@20 on
the x-axis) improves. Notably, the strategy of directly replacing cold
item weights can sustain cold item recommendation performance without
significantly harming overall accuracy.}

\end{figure}

We further examine the effect of hyperparameters on the two variants of
the FEASE model, either optimized jointly with the prior similarity
matrix \(R\) (i.e.~models with -Prior suffix,
Equation~\ref{eq-fease-i-prior}) or simply replacing the cold-item
weights using this matrix (i.e.~models without -Prior suffix,
Equation~\ref{eq-fease-i-prior-masked} or the equivalent heuristic).
Varying the prior weight hyperparameter \(\delta\) results a tradeoff
between warm and cold item recommendation.
Figure~\ref{fig-joint-vs-replacement} plots the HitRatio@20 metric
against the ColdItemHR@20 on models trained on the Netflix dataset,
effectively comparing the compromise between warm and cold item
recommendation accuracy. We observe that, to achieve a certain level of
improvement in handling item cold start issues, the joint optimization
models tend to sacrifice more in warm item recommendation compared to
the simple weight replacement models. However, we also observe that it
is possible to tune the jointly optimized models to achieve even higher
accuracy than simple weight replacement models, though completely
sacrificing the ability to handle item cold start. Importantly, since
the capability of handling item cold start is governed by \(\delta\), we
essentially have a knob that allows us to fine-tune the models to
exhibit different levels of cold start handling effectiveness. In
practice, we may create different user experiences in different parts of
the application, using identical modeling approach.

\section{Conclusion}\label{sec-conclusion}

In this work, we presented the FEASE model---a unified framework
designed to address both item and user cold start challenges by
effectively incorporating relevant side information. We examined the
inherent balance between recommending for warm users and items vs.~those
facing cold start issues within collaborative filtering systems. Similar
to the EASE model, our approach benefits from a closed-form solution,
making it straightforward to implement in Python. Looking ahead, future
research could explore ways to further refine this model, such as by
integrating additional forms of side information (e.g., textual, visual,
or contextual data), adapting the approach to dynamic or real-time
recommendation environments, or combining it with deep learning
techniques to better capture complex user-item interactions. Moreover,
evaluating its performance across various application domains could
offer valuable insights into its versatility and scalability.

\section{Acknowledgement}\label{acknowledgement}

The manuscript is written with the help of OpenAI's ChatGPT for grammar
and style refinement. All ideas of the paper are entirely the work of
the authors.

\bibliographystyle{ACM-Reference-Format}
\bibliography{references.bib}


\begin{thebibliography}{17}


\ifx \showCODEN    \undefined \def \showCODEN     #1{\unskip}     \fi
\ifx \showDOI      \undefined \def \showDOI       #1{#1}\fi
\ifx \showISBNx    \undefined \def \showISBNx     #1{\unskip}     \fi
\ifx \showISBNxiii \undefined \def \showISBNxiii  #1{\unskip}     \fi
\ifx \showISSN     \undefined \def \showISSN      #1{\unskip}     \fi
\ifx \showLCCN     \undefined \def \showLCCN      #1{\unskip}     \fi
\ifx \shownote     \undefined \def \shownote      #1{#1}          \fi
\ifx \showarticletitle \undefined \def \showarticletitle #1{#1}   \fi
\ifx \showURL      \undefined \def \showURL       {\relax}        \fi
\providecommand\bibfield[2]{#2}
\providecommand\bibinfo[2]{#2}
\providecommand\natexlab[1]{#1}
\providecommand\showeprint[2][]{arXiv:#2}

\bibitem[Deshpande and Karypis(2004)]%
        {deshpandeItembasedTopRecommendation2004a}
\bibfield{author}{\bibinfo{person}{Mukund Deshpande} {and}
  \bibinfo{person}{George Karypis}.} \bibinfo{year}{2004}\natexlab{}.
\newblock \showarticletitle{Item-Based Top- {{N}} Recommendation Algorithms}.
\newblock \bibinfo{journal}{\emph{ACM Transactions on Information Systems -
  TOIS}}  \bibinfo{volume}{22} (\bibinfo{date}{Jan.} \bibinfo{year}{2004}),
  \bibinfo{pages}{143--177}.
\newblock
\urldef\tempurl%
\url{https://doi.org/10.1145/963770.963776}
\showDOI{\tempurl}


\bibitem[{Gomez-Uribe} and Hunt(2015)]%
        {gomez-uribeNetflixRecommenderSystem2015}
\bibfield{author}{\bibinfo{person}{Carlos~A. {Gomez-Uribe}} {and}
  \bibinfo{person}{Neil Hunt}.} \bibinfo{year}{2015}\natexlab{}.
\newblock \showarticletitle{The Netflix Recommender System: {{Algorithms}},
  Business Value, and Innovation}.
\newblock \bibinfo{journal}{\emph{ACM Transactions on Management Information
  Systems}} \bibinfo{volume}{6}, \bibinfo{number}{4} (\bibinfo{year}{2015}).
\newblock
\showISSN{21586578}
\urldef\tempurl%
\url{https://doi.org/10.1145/2843948}
\showDOI{\tempurl}


\bibitem[Harper and Konstan(2016)]%
        {harperMovieLensDatasetsHistory2016}
\bibfield{author}{\bibinfo{person}{F.~Maxwell Harper} {and}
  \bibinfo{person}{Joseph~A. Konstan}.} \bibinfo{year}{2016}\natexlab{}.
\newblock \showarticletitle{The {{MovieLens Datasets}}: {{History}} and
  {{Context}}}.
\newblock \bibinfo{journal}{\emph{ACM Transactions on Interactive Intelligent
  Systems}} \bibinfo{volume}{5}, \bibinfo{number}{4} (\bibinfo{date}{Jan.}
  \bibinfo{year}{2016}), \bibinfo{pages}{1--19}.
\newblock
\showISSN{2160-6455, 2160-6463}
\urldef\tempurl%
\url{https://doi.org/10.1145/2827872}
\showDOI{\tempurl}


\bibitem[Harris et~al\mbox{.}(2020)]%
        {harrisArrayProgrammingNumPy2020}
\bibfield{author}{\bibinfo{person}{Charles~R. Harris},
  \bibinfo{person}{K.~Jarrod Millman}, \bibinfo{person}{St{\'e}fan~J. {van der
  Walt}}, \bibinfo{person}{Ralf Gommers}, \bibinfo{person}{Pauli Virtanen},
  \bibinfo{person}{David Cournapeau}, \bibinfo{person}{Eric Wieser},
  \bibinfo{person}{Julian Taylor}, \bibinfo{person}{Sebastian Berg},
  \bibinfo{person}{Nathaniel~J. Smith}, \bibinfo{person}{Robert Kern},
  \bibinfo{person}{Matti Picus}, \bibinfo{person}{Stephan Hoyer},
  \bibinfo{person}{Marten~H. {van Kerkwijk}}, \bibinfo{person}{Matthew Brett},
  \bibinfo{person}{Allan Haldane}, \bibinfo{person}{Jaime~Fern{\'a}ndez {del
  R{\'i}o}}, \bibinfo{person}{Mark Wiebe}, \bibinfo{person}{Pearu Peterson},
  \bibinfo{person}{Pierre {G{\'e}rard-Marchant}}, \bibinfo{person}{Kevin
  Sheppard}, \bibinfo{person}{Tyler Reddy}, \bibinfo{person}{Warren Weckesser},
  \bibinfo{person}{Hameer Abbasi}, \bibinfo{person}{Christoph Gohlke}, {and}
  \bibinfo{person}{Travis~E. Oliphant}.} \bibinfo{year}{2020}\natexlab{}.
\newblock \showarticletitle{Array Programming with {{NumPy}}}.
\newblock \bibinfo{journal}{\emph{Nature}} \bibinfo{volume}{585},
  \bibinfo{number}{7825} (\bibinfo{date}{Sept.} \bibinfo{year}{2020}),
  \bibinfo{pages}{357--362}.
\newblock
\showISSN{1476-4687}
\urldef\tempurl%
\url{https://doi.org/10.1038/s41586-020-2649-2}
\showDOI{\tempurl}


\bibitem[Jeunen et~al\mbox{.}(2020)]%
        {jeunenClosedFormModelsCollaborative2020}
\bibfield{author}{\bibinfo{person}{Olivier Jeunen}, \bibinfo{person}{Jan
  Van~Balen}, {and} \bibinfo{person}{Bart Goethals}.}
  \bibinfo{year}{2020}\natexlab{}.
\newblock \showarticletitle{Closed-{{Form Models}} for {{Collaborative
  Filtering}} with {{Side-Information}}}. In
  \bibinfo{booktitle}{\emph{Fourteenth {{ACM Conference}} on {{Recommender
  Systems}}}}. \bibinfo{publisher}{ACM}, \bibinfo{address}{Virtual Event
  Brazil}, \bibinfo{pages}{651--656}.
\newblock
\showISBNx{978-1-4503-7583-2}
\urldef\tempurl%
\url{https://doi.org/10.1145/3383313.3418480}
\showDOI{\tempurl}


\bibitem[Liang et~al\mbox{.}(2018)]%
        {liangVariationalAutoencodersCollaborative2018}
\bibfield{author}{\bibinfo{person}{Dawen Liang}, \bibinfo{person}{Rahul~G.
  Krishnan}, \bibinfo{person}{Matthew~D. Hoffman}, {and} \bibinfo{person}{Tony
  Jebara}.} \bibinfo{year}{2018}\natexlab{}.
\newblock \showarticletitle{Variational Autoencoders for Collaborative
  Filtering}.
\newblock \bibinfo{journal}{\emph{The Web Conference 2018 - Proceedings of the
  World Wide Web Conference, WWW 2018}} (\bibinfo{year}{2018}),
  \bibinfo{pages}{689--698}.
\newblock
\showISBNx{9781450356398}
\urldef\tempurl%
\url{https://doi.org/10.1145/3178876.3186150}
\showDOI{\tempurl}
\showeprint[arxiv]{1802.05814}


\bibitem[Monteil et~al\mbox{.}(2024)]%
        {monteilMARecMetadataAlignment2024a}
\bibfield{author}{\bibinfo{person}{Julien Monteil}, \bibinfo{person}{Volodymyr
  Vaskovych}, \bibinfo{person}{Wentao Lu}, \bibinfo{person}{Anirban Majumder},
  {and} \bibinfo{person}{Anton Van Den~Hengel}.}
  \bibinfo{year}{2024}\natexlab{}.
\newblock \showarticletitle{{{MARec}}: {{Metadata Alignment}} for Cold-Start
  {{Recommendation}}}. In \bibinfo{booktitle}{\emph{18th {{ACM Conference}} on
  {{Recommender Systems}}}}. \bibinfo{publisher}{ACM}, \bibinfo{address}{Bari
  Italy}, \bibinfo{pages}{401--410}.
\newblock
\showISBNx{9798400705052}
\urldef\tempurl%
\url{https://doi.org/10.1145/3640457.3688125}
\showDOI{\tempurl}


\bibitem[{Netflix}(2006)]%
        {netflixNetflixPrizeData2006}
\bibfield{author}{\bibinfo{person}{{Netflix}}.}
  \bibinfo{year}{2006}\natexlab{}.
\newblock \bibinfo{title}{Netflix {{Prize Data}}}.
\newblock
  \bibinfo{howpublished}{https://www.kaggle.com/datasets/netflix-inc/netflix-prize-data}.
\newblock


\bibitem[Ni et~al\mbox{.}(2019)]%
        {niJustifyingRecommendationsUsing2019}
\bibfield{author}{\bibinfo{person}{Jianmo Ni}, \bibinfo{person}{Jiacheng Li},
  {and} \bibinfo{person}{Julian McAuley}.} \bibinfo{year}{2019}\natexlab{}.
\newblock \showarticletitle{Justifying {{Recommendations}} Using
  {{Distantly-Labeled Reviews}} and {{Fine-Grained Aspects}}}. In
  \bibinfo{booktitle}{\emph{Proceedings of the 2019 {{Conference}} on
  {{Empirical Methods}} in {{Natural Language Processing}} and the 9th
  {{International Joint Conference}} on {{Natural Language Processing}}
  ({{EMNLP-IJCNLP}})}}, \bibfield{editor}{\bibinfo{person}{Kentaro Inui},
  \bibinfo{person}{Jing Jiang}, \bibinfo{person}{Vincent Ng}, {and}
  \bibinfo{person}{Xiaojun Wan}} (Eds.). \bibinfo{publisher}{Association for
  Computational Linguistics}, \bibinfo{address}{Hong Kong, China},
  \bibinfo{pages}{188--197}.
\newblock
\urldef\tempurl%
\url{https://doi.org/10.18653/v1/D19-1018}
\showDOI{\tempurl}


\bibitem[Ning and Karypis(2011)]%
        {ningSLIMSparseLinear2011}
\bibfield{author}{\bibinfo{person}{Xia Ning} {and} \bibinfo{person}{George
  Karypis}.} \bibinfo{year}{2011}\natexlab{}.
\newblock \showarticletitle{{{SLIM}} : {{Sparse Linear Methods}} for {{Top-N
  Recommender Systems}}}.
\newblock  (\bibinfo{date}{Dec.} \bibinfo{year}{2011}), \bibinfo{pages}{1--10}.
\newblock


\bibitem[Ning and Karypis(2012)]%
        {ningSparseLinearMethods2012}
\bibfield{author}{\bibinfo{person}{Xia Ning} {and} \bibinfo{person}{George
  Karypis}.} \bibinfo{year}{2012}\natexlab{}.
\newblock \showarticletitle{Sparse Linear Methods with Side Information for
  {{Top-N}} Recommendations}. \bibinfo{pages}{581--582}.
\newblock
\urldef\tempurl%
\url{https://doi.org/10.1145/2187980.2188137}
\showDOI{\tempurl}


\bibitem[Rendle(2010)]%
        {rendleFactorizationMachines2010}
\bibfield{author}{\bibinfo{person}{Steffen Rendle}.}
  \bibinfo{year}{2010}\natexlab{}.
\newblock \showarticletitle{Factorization Machines}.
\newblock \bibinfo{journal}{\emph{Proceedings - IEEE International Conference
  on Data Mining, ICDM}} (\bibinfo{year}{2010}), \bibinfo{pages}{995--1000}.
\newblock
\showISBNx{9780769542560}
\showISSN{15504786}
\urldef\tempurl%
\url{https://doi.org/10.1109/ICDM.2010.127}
\showDOI{\tempurl}


\bibitem[Rendle et~al\mbox{.}(2009)]%
        {rendleBPRBayesianPersonalized2009}
\bibfield{author}{\bibinfo{person}{Steffen Rendle}, \bibinfo{person}{Christoph
  Freudenthaler}, \bibinfo{person}{Zeno Gantner}, {and} \bibinfo{person}{Lars
  {Schmidt-thieme}}.} \bibinfo{year}{2009}\natexlab{}.
\newblock \showarticletitle{{{BPR}} : {{Bayesian Personalized Ranking}} from
  {{Implicit Feedback}}}.
\newblock  (\bibinfo{year}{2009}), \bibinfo{pages}{452--461}.
\newblock


\bibitem[Sedhain et~al\mbox{.}(2015)]%
        {sedhainAutoRecAutoencodersMeet2015}
\bibfield{author}{\bibinfo{person}{Suvash Sedhain},
  \bibinfo{person}{Aditya~Krishna Menon}, \bibinfo{person}{Scott Sanner}, {and}
  \bibinfo{person}{Lexing Xie}.} \bibinfo{year}{2015}\natexlab{}.
\newblock \showarticletitle{{{AutoRec}} : {{Autoencoders Meet Collaborative
  Filtering}}}.
\newblock  (\bibinfo{year}{2015}), \bibinfo{pages}{0--1}.
\newblock
\showISBNx{9781450334730}


\bibitem[Steck(2019)]%
        {steckEmbarrassinglyShallowAutoencoders2019}
\bibfield{author}{\bibinfo{person}{Harald Steck}.}
  \bibinfo{year}{2019}\natexlab{}.
\newblock \showarticletitle{Embarrassingly {{Shallow Autoencoders}} for
  {{Sparse Data}}}.
\newblock  (\bibinfo{date}{May} \bibinfo{year}{2019}).
\newblock
\showeprint[arxiv]{1905.03375v1}


\bibitem[Volkovs et~al\mbox{.}(2017)]%
        {volkovsDropoutNetAddressingCold2017}
\bibfield{author}{\bibinfo{person}{Maksims Volkovs}, \bibinfo{person}{Guangwei
  Yu}, {and} \bibinfo{person}{Tomi Poutanen}.} \bibinfo{year}{2017}\natexlab{}.
\newblock \showarticletitle{{{DropoutNet}}: {{Addressing Cold Start}} in
  {{Recommender Systems}}}. In \bibinfo{booktitle}{\emph{Advances in {{Neural
  Information Processing Systems}}}}, Vol.~\bibinfo{volume}{30}.
  \bibinfo{publisher}{Curran Associates, Inc.}
\newblock


\bibitem[Zhao et~al\mbox{.}(2016)]%
        {zhaoCollaborativeDeepDenoising2016}
\bibfield{author}{\bibinfo{person}{Jinjin Zhao}, \bibinfo{person}{Lei Wang},
  \bibinfo{person}{Dong Xiang}, {and} \bibinfo{person}{Brett Johanson}.}
  \bibinfo{year}{2016}\natexlab{}.
\newblock \showarticletitle{Collaborative {{Deep Denoising Autoencoder
  Framework}} for {{Recommendations}}}.
\newblock  (\bibinfo{date}{Feb.} \bibinfo{year}{2016}).
\newblock


\end{thebibliography}


\end{document}